\documentclass{article}
\usepackage{fleqn,espcrc2, epsfig}

\def\fun#1#2{\lower3.6pt\vbox{\baselineskip0pt\lineskip.9pt
\ialign{$\mathsurround=0pt#1\hfil##\hfil$\crcr#2\crcr\sim\crcr}}}
\newcommand{\be}{\begin{equation}}
\newcommand{\ee}{\end{equation}}
\newcommand{\ba}{\begin{eqnarray}}
\newcommand{\ea}{\end{eqnarray}}
\newcommand{\bc}{\begin{center}}
\newcommand{\ec}{\end{center}}
\begin{document}

\title{\noindent The bound state corrections to the semileptonic decays of the
heavy baryons}
\author{I. D'Souza$^{(a)}$, C. S. Kalman$^{(a)},$P. Yu
Kulikov$^{(b){\ }}{and}$ I.
M. Narodetskii$^{(b){\ }}$ \\
$^{(a)}$Concordia University, Montreal, Canada \\
$^{(b)}$Institute of Theoretical and Experimental Physics, Moscow,
 RF}

\begin{abstract}
We present an investigation of the lepton energy distributions in the
inclusive semileptonic weak decays of heavy baryons performed within a
relativistic quark model formulated on the light front (LF).
Using the heavy-quark LF distribution functions related to the equal time
momentum wave functions taken from the Kalman--Tran--D'Souza model
we compute the electron energy spectra and the total semileptonic widths of
the $\Lambda _{Q}$ and $\Xi _{Q}$ ($Q=c,b$) baryons and confront the results
with existing data.

\end{abstract}

\maketitle

\section{Introduction}

It is important to study the differential distributions in
semileptonic decays of heavy flavors in order to extract the CKM
matrix elements$|V_{cb}|$ and $|V_{ub}|$, key phenomenological
parameters of the Standard Model.
To this end one needs to disentagle the effects of the strong
interactions at large distances from the quark-gluon Lagrangian
known at small distances. Since the $b$--quark is heavy compared
to the QCD scale, the inclusive semileptonic decays can be treated
with the help of an operator product expansion (OPE) combined with
the heavy quark expansion (HQE) \cite{BSU97}. The result (away
from the endpoint of the spectrum) is that the inclusive
differential decay width $d\Gamma /dE_{\ell }$ ( $E_{\ell }$ is
the lepton energy) may be expanded in $\Lambda /m_{b}$, where
$\Lambda $ is a QCD related scale of order 500 MeV, and $m_{b}$ is
the mass of the heavy quark. The leading term (zeroth order in
$\Lambda /m_{b}$) is the free quark decay spectrum, the subleading
term vanishes, and the subsubleading term involves parameters from
the heavy quark theory, but should be rather small, as it is of
order $(\Lambda /m_{b})^{2}$. However, near the end point the
$1/m_{b}$ expansion has to be replaced by an expansion in twist.
To describe this region one has to introduce a so--called ``shape
function'', which in
principle introduces a large hadronic uncertainty. 
Since a model independent determination of the shape function is not
available at the present time,  a certain model dependence in this region
seems to be unavoidable.

The global prediction of OPE is that the lepton spectra from the
semileptonic decays of the heavy flavor hadrons are all alike with
accuracy up to ${\cal O}(1/m_b^{3})$ except for the end point
region. However, it is well known that the similar OPE prediction
for the $\Lambda _{b}$ lifetime is puzzling because it predicts a
value of $\tau _{B}/\tau _{\Lambda _{B}}$ , which is considerably
smaller than determined experimentally. It is not clear whether
the present contradiction between the theory and the data is a
temporary difficulty, or whether it implies a fundamental flaw in
the OPE approach. In this respect, the use of phenomenological
models, like the constituent quark model, could be of great
interest as a complementary approach to the $OPE$ method.
\section{The light-front model}
 Two phenomenological approaches had been applied to
describe strong interaction effects in the inclusive weak decays:
the parton ACM model amended to include the motion of the heavy
quark inside the decaying hadron \cite{ACM82}, and the ``exclusive
model'' based on the summation of different channels, one by one
\cite{ISGW89}. 
The various light--front (LF) approaches to consideration of the inclusive
semileptonic transitions were suggested in Refs. \cite{JP94}--\cite{KNST99}.
Each of them needs an input from constituent quark model to parametrize
non-perturbative effects.

In this paper, we evaluate the non--perturbative corrections to
the lepton spectrum in the
inclusive semileptonic decays of heavy flavor baryons
using an approach of Ref. \cite{KNST99} in which non-perturbative
$QCD$ effects in a heavy baryon are mocked up by a constituent
quark model wave function $\psi (\xi ,p_{\bot }^{2})$ on the LF.
$|\psi (\xi ,p_{\bot }^{2})|^{2}$ represents the probability to
find the $Q$ quark\footnote{Here and below the symbol $Q=b,c$
denotes the heavy quark and the symbol $H_Q$ denotes the heavy
hadron.}  carrying a LF fraction $\xi $ and a transverse momentum
squared $p_{\bot }^{2}=|{\bf p}_{\bot }|^{2}$. The calculations of
the inclusive semileptonic widths and the corresponding branching
ratios found here introduce no new parameters. It is based solely
upon the eigenfunctions for baryon states found by Kalman, Tran
and D'Souza (see \cite {DKT}). Our main purpose here is to
confront the lepton spectra from semileptonic decays of the
different baryons.


The general expression for the distribution over lepton energy is
based on the standard Lorentz--invariant kinematical analysis (see
{\it e.g.} \cite{BKSV93}):

\begin{eqnarray}
\frac{d\Gamma _{SL}}{dy} &=&\frac{G_{F}^{2}m_{H_Q}^{5}}{64\pi ^{3}}%
|V_{Q'Q}|^{2}\int\limits_{0}^{t_{max}}dt\int\limits_{s_{0}}^{s_{max}}ds
 \\ &&\left\{
\begin{array}{ll}
 & tW_{1}+\frac{1}{2}[y(1+t-s)-y^{2}-t]W_{2}\nonumber\\[3mm]
 & +t[\frac{1+t-s}{2}-y]W_{3}+\ldots
\end{array}
\right\}   \label{13}
\end{eqnarray}
where $y=2E_{\ell }/m_{H_Q}$, $t=q^{2}/m_{H_Q}^{2}$,
$s=M_{X_{Q'}}^{2}/m_{H_Q}^{2}$, $ q=p_{\ell }+p_{\nu _{\ell }}$,
$M_{X_Q'}^{2}=(p_{H_Q}-q)^{2}$ and the structure functions
$W_{i}=W_{i}(s,t)$ appear in the decomposition of the hadronic
tensor $W_{\alpha \beta }$ in Lorentz covariants. The ellipsis in
(\ref{13}) denotes the terms proportional to the lepton mass
squared. The kinematical limits of integration can be found from
equation
\begin{equation}
\frac{s}{1-y}+\frac{t}{y}\leq 1.  \label{14}
\end{equation}
They are given by $ 0\leq y\leq y_{max}$, $s_{max}=1+t-(y+t/y)$,
and $t_{max}=y((1-(1-y_{max})/(1-y)),$ with $y_{max}
=1-(m_{X_{Q'}^{(min)}}/m_{H_Q})^{2}$ and $m_{X_{Q'}}^{(min)}$
being the minimal mass of the hadronic state $X_{Q'}$.

In a parton model of Ref. \cite{KNST99} inclusive semileptonic
$H_Q\to X_Q'\ell \nu _{\ell }$ decay is treated in a direct
analogy to deep-inelastic scattering. Specifically, it is assumed
that the sum over all possible  final states $X_{Q'}$ can be
modelled by the decay width of an on--shell $Q$ quark into
on--shell $Q'$--quark weighted with the $Q$--quark distribution
$|\psi(\xi ,p_{\bot }^{2})|^2$. The natural variables for
$\psi(\xi ,p_{\bot }^{2})$ are the light-cone momentum fraction
$\xi =p_{Q}^{+}/(p_{1}^{+}+p_{2}^{+}+p_{Q}^{+})$ and the $Q$ quark
transverse momentum $p_{\bot }^{2}={\bf p}_{Q\bot }$. The function
$|\psi(\xi ,p_{\bot }^{2})|^2$ has support $0\le \xi \le 1$ and
$0\le p_{\bot }^{2}\le \infty $. This function as a function of
$\xi $ has the maximum at $\xi =m_{Q}/(m_{Q}+m_1+m_2)$, at smaller
values of $\xi $ it rapidly vanishes. The normalization condition
reads
\begin{equation}
\pi \int\limits_{0}^{1}d\xi \int dp_{\bot }^{2}|\psi(\xi ,p_{\bot
}^{2})|^2=1. \label{normalization_condition}
\end{equation}
Following the above assumption, the hadronic tensor $W_{\alpha \beta }$ is
written as
\begin{eqnarray}
W_{\alpha \beta }&=&\int w_{\alpha \beta
}^{(Q'Q)}(p_{Q'},p_{Q})\delta
[(p_{Q}-q)^{2}-m_{Q'}^{2}]\nonumber\\
&&\frac{|\psi(\xi ,p_{\bot }^{2})|^2}{\xi }\theta (\varepsilon
_{Q'})d\xi d^{2}p_{\bot }, \label{w_alpha_beta}
\end{eqnarray}
where
\begin{equation}
w_{\alpha \beta
}^{(Q'Q)}(p_{Q'},p_{Q})=\frac{1}{2}\sum_{spins}\bar{u}
_{Q'}O_{\alpha }u_{Q}\cdot \bar{u}_{Q}O_{\beta }^{+}u_{Q'},
\label{17}
\end{equation}
with $O_{\alpha }=\gamma _{\alpha }(1-\gamma _{5})$. The factor
$1/\xi$ in Eq. (\ref{w_alpha_beta}) is due to the normalization of
the hadron-quark vertex \cite{DGNS97}. The hadronic tensor
(\ref{w_alpha_beta}) differs from the corresponding expressions of
Refs. \cite{JP94} and \cite{MTM96} by the non--trivial dependence
on $p_{\bot }^{2}$ which enters both $|\psi (\xi ,p_{\bot
}^{2})|^{2}$ and argument of the $\delta $--function. The
calculation of the structure functions $W_{i}(t,s)$ from Eq.
(\ref{w_alpha_beta}) is straightforward. For details see Appendix
of \cite{KNST99}.

Note that in the heavy-quark limit, $m_{Q}\to \infty $, the
$Q$-quark distribution in a heavy hadron becomes a delta function
peaked at $\xi=1$,
which implies that the semileptonic width of a heavy hadrons
coincides with that of a heavy quark. At finite values of the
$Q$-quark mass the result for $\Gamma _{SL}$ exhibits in general
an $m_{Q}$-dependence of the following general form: $\Gamma
_{SL}\propto m_{Q}^{5}[1+c/m_{Q}+O(1/m_{Q}^{2})]$, where $c$ is a
non-vanishing coefficient depending on the particular quark model
adopted. However, the mass $m_{Q}$ is the constituent mass of the
$b$-quark, which may differ from the pole quark mass $m_{Q}^{{\rm
pole}}$ commonly appearing in the $OPE$ of the heavy hadron decay
rate (see below). Assuming $m_{Q}=m_{Q}^{pole}-c/5+O(1/m_{Q}^{
{\rm pole}})$, the well-known result \cite{BSU97} of the absence of the
$
1/\mu _{Q}$ corrections to the free-quark decay may be recovered.
The above argument is completely analogous to that used to
eliminate the $1/m_{b}$ corrections from the total width in the
ACM model \cite{BSUV94}.
\section{The electron spectrum from semileptonic {\cal B} decays}
Before calculating the electron spectra for heavy baryons we
illustrate our approach with an example of  the electron spectrum
from the inclusive semileptonic {\cal B} decays \cite{GKN00}.

Since we do not have an explicit representation for the {\cal
B}--meson Fock expansion in QCD, we proceed by making an ans\"atz
for $\psi (\xi ,p_{\bot }^{2})$. This is a model dependent
enterprise but has its close equivalent in studies of electron
spectra using the ACM model. We first make an ans\"atze for the
momentum space structure of an equal time (ET) quark model wave
function $\Phi({\bf p}^2)$ and then convert from ET to LF momenta
by leaving the transverse momenta unchanged and letting \be
\label{change_of_variables}
p_{iz}=\frac{1}{2}(p_i^+-p_i^-)=\frac{1}{2}(p_i^+-\frac{p^2_{i{\bot}}
+m^2_i}{p_i^+}) \ee for both the $b$--quark $(i=b)$ and the
quark--spectator $(i=sp)$. Because the ET function depends on the
relative momentum it is more convenient to use the
quark--antiquark rest frame instead of the $\cal B$--meson rest
frame. Recall that in the LF formalism these two frames are
different. Then the longitudinal LF momentum fractions $\xi_i$ are
defined as $\xi_{sp}=p^+_{sp}/M_0$, $\xi_b=p^+_b/M_0$, where the
free mass operator $M_0^2$ is defined as \begin{eqnarray}
\label{M_0}
M_0^2&=&\left(\sqrt{m^2_Q+{\bf p}^2}+\sqrt{m^2_{sp}+{\bf p}^2}\right)^2=\nonumber\\
&&=\frac{p^2_{\perp}+m^2_b}{\xi}+\frac{p^2_{\perp}+m^2_{sp}}{1-\xi},
\end{eqnarray} with  ${\bf
p}^2=p^2_{\bot}+p^2_z$, and \be
p_z=(x-\frac{1}{2})M_0-\frac{m^2_b-m^2_{sp}}{2M_0}.\ee

The distribution function $|\psi (\xi ,p_{\bot }^{2})|^{2}~(\xi
=\xi_{b})$ normalized according to (\ref
{normalization_condition}) is given by
$|\psi(\xi,p^2_{\bot})|^2=|\partial p_z/\partial
\xi|^2\Phi(p^2_{\bot}+p^2_z(\xi,p^2_{\bot})). $ Explicit form of
$|\partial p_z/\partial \xi|$ is given {\it e.g.} by Eq. (22) of
Ref. \cite{KNST99}. In this calculation the equal time momentum
distribution $\Phi ({\bf p}^{2})$ of a quark-spectator is taken in
the standard Gaussian form \be \label{gauss} \Phi({\bf
p}^2)=\frac{4}{\sqrt{\pi} p_F^3}\exp\left(-\frac{{\bf
p}^2}{p_F^2}\right), \ee with the Fermi momentum $p_{F}=0.4$ GeV.
The calculation uses $m_{b}=4.8$ GeV (the average value of the
floating $b$-quark mass in the ACM model) and $m_{c}=1.5$ GeV.

We have implicitly included the ${\cal O}(\alpha _{s})$
perturbative corrections with $\alpha _{s}=0.25$ arising from
gluon Bremsstrahllung and one--loop effects which modify an
electron energy spectra at the partonic level. It is customary to
define a correction function $G(x)$ to the electron spectrum
$d\Gamma _{b}^{(0)}$ calculated in the tree approximation for the
free quark decay through
\begin{equation}
\frac{d\Gamma _{b}}{dx}=\frac{d\Gamma _{b}^{(0)}}{dx}\left(
1-\frac{2\alpha _{s}}{3\pi }G(x)\right) ,
\end{equation}
where $x=2E/m_{b}$.
In
actual calculations we neglect the terms $\sim m_c^2/m_b^2 $ in
$G(x) $ and take this function from \cite{JK89}.
\begin{figure}
\begin{center}
\epsfxsize=7.5cm \epsfysize=7.5cm \epsfbox{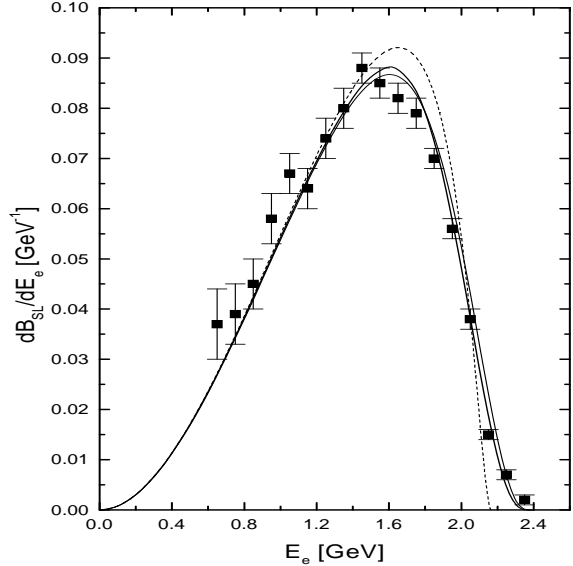}
\end{center}
\vskip -1cm \caption{The predicted electron energy spectrum the
semileptonic $\cal B$ decays compared with the CLEO data
\cite{CLEO}. Thick solid line is the LF result, thin solid line is
the ACM result, dashed line refers to the free quark decay.
}
\end{figure}

 We display
in Fig. 1 the results of our calculations of the electron spectrum
in the inclusive semileptonic ${\cal B}$ decays.  We find an
overall agreement with the published experimental data from the
CLEO collaboration \cite{CLEO}. A more detailed fit to the
measured spectrum can impose constraints on the distribution
function and the mass of the charm quark. Such the fit should also
account for detector resolution. The LF and ACM spectra are
normalized to $10.71\%$, $10.84\%$,
respectively. $|V_{cb}|=0.042$. Using $\tau _{B}=1.653\pm 0.028$
ps$^{-1}$ \cite{pdg00} one obtains
$\Gamma ({\cal B}\to X_{c}e\bar{\nu})=(6.51\pm 0.11) \times
10^{-2}~ ps^{-1}~~({\rm LF}),$
and
$\Gamma ({\cal B}\to X_{c}e\bar{\nu} )=(6.56\pm 0.11)\times
10^{-2}~ ps^{-1}~({\rm ACM})$,
that agrees within the error bars with the experimental
result
$\Gamma _{exp}(b \to X_{c}e\bar{\nu})=(6.67\pm 0.2)\times
10^{-2}~{\rm ps}^{-1} $
(for
$|V_{cb}|_{incl}=(40.4\pm 1.07)\times 10^{-3}$), quoted in the
forthcoming PDG2002 \cite{pdg00}
\footnote{%
It should be noted that $\Gamma _{exp}$ in \cite{pdg00} involves a
weighted
average of the ${\cal B}$ mesons and $\Lambda _{b}$ baryons produced in $%
Z^{0}$ decays (hence the symbol $b$) different from the
corresponding one given by CLEO, which has been measured at the
$\Upsilon (4S)$ resonance.}.


\section{The electron spectrum for the heavy baryon semileptonic
decays} We proceed to present numerical results for the lepton
spectra in the heavy baryon decays. We consider the problem of a
constituent bound state formed by a heavy quark interacting with
the lighter ones. The spectator quarks have a momentum
distribution $\Phi ({\bf p}_{1},{\bf p}_{2})$ where ${\bf p}_{1}$
and ${\bf p}_{2}$ are their 3-momenta. Specifically, in the
Kalman--Tran-D'Souza model in the first approximation the momentum
distribution of spectator quarks is: \vspace{2mm}

\begin{eqnarray}
\Phi ({\bf p}_{1},{\bf p}_{2}) &=&(\frac{1}{\alpha\sqrt{\pi
}})^{3}(
\frac{1}{\beta\sqrt{\pi }})^{3}  \nonumber \\
&&\times \exp (-\frac{{\bf k}^{2}}{\alpha ^{2}})\exp (- \frac{{\bf
p}_Q^2}{\beta^{2}}), \label{gaussian_distribution}
\end{eqnarray}
\vspace{2mm}

\noindent where ${\bf k}=(m_2{\bf p}_{1}-m_1{\bf
p}_{2})/(m_1+m_2)$, $m_1$ and $m_2$ being the masses of the
spectator quarks. The parameters $\alpha$ and $\beta$ were are
obtained from the fit to the spectra in Refs. \cite{DKT}. These
parameters are listed in Table 2 of Ref. \cite{DKKN00} where the
comparison of the theoretical predictions of the Kalman-D'Souza
model for the heavy baryon masses with the experimental data was
also given. The heavy quark distribution function is given by
\begin{eqnarray}
\label{F}
 F(\xi ,{\bf p}_{Q\bot })
&=&\frac{1}{8}\int\limits_{0}^{1-\xi }d\eta
\int
d{\bf k}_{\bot}\Phi({\bf p}_1,{\bf p}_2)\nonumber\\
&&\times\left( M_{0}+\frac{
e_Q}{\xi M_0}\right)
\left( M_{0}+\frac{e_1}{\eta M_{0}}\right)   \nonumber \\
&&\times \left( M_{0}+\frac{e_2}{( 1-\xi -\eta)M_{0}}\right)
\end{eqnarray}
where $\eta=p_1^+/M_0$, $e_Q = (p^2_{Q\bot}+m^2_Q)/\xi$, $
e_1=(k^2_{\bot}+m_1^{2})/\eta$, $e_2=\left(({\bf k}_{\bot}-{\bf
p}_{Q\bot})^2+m_2^{2}\right)/(1-\xi-\eta)$,
 and the free mass operator
$M_0^2$ is now given by $M_0^2=e_Q+e_1+e_2$. In terms of the LF
variables ${\bf k}^2$ and ${\bf p}_Q^2$ are given by
\begin{eqnarray}
{\bf k}^2&=&{\bf k}_{\bot}^2-\frac{2m_1}{m_1+m_2}{\bf
k}_{\bot}{\bf p}_{Q\bot} +\left(\frac{m_1}{m_1+m_2}\right)^2{\bf
p}_{Q\bot}^2\nonumber\\
&&+\frac{M_0^2}{4}\times\left(\eta+\frac{m_1}{m_1+m_2}\xi\right.\nonumber\\
&&\left.-\frac{m_1}{m_1+m_2}\cdot\frac{e_Q}{M_0^2}-\frac{e_1}{M_0^2}\right)^2,
\end{eqnarray}
and
\begin{equation}
{\bf p}_{Q}^{2}={\bf p}_{Q\bot }^{2}+\frac{1}{4}\left( { \xi
M}_{0}-\frac{e_Q}{{M}_{0}}\right) ^{2}. \label{24}
\end{equation}

\begin{figure}
\begin{center}
\epsfxsize=7.5cm \epsfysize=7.5cm \epsfbox{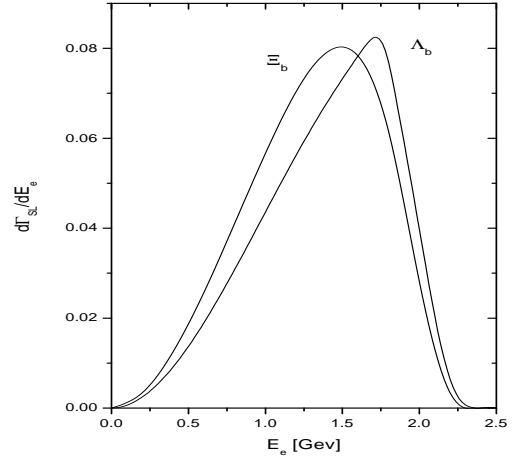}
\end{center}
\vskip -1cm \caption{The predicted electron energy spectrum for $%
\Lambda_b$ and $\Xi_b$ semileptonic decays.  The spectra
normalized to $\Gamma_{SL}(\Lambda_b)=0.086$ ps$^{-1}$ and
$\Gamma_{SL}(\Xi_b)=0.092$ ps$^{-1}$. $|V_{cb}|=0.04$}
\end{figure}

In the calculation of the lepton spectra instead of the
identification $ m_{Q}^{LF}=<m_{f}^{ACM}>$ used in the ${\cal B}$
meson case before we use the original values of the quark masses
\cite{DKT}
\begin{eqnarray}
m_{u} &=&m_{d}=0.23~{\rm GeV},~~m_{s}=0.605~{\rm GeV},~  \nonumber \\
~m_{c} &=&1.961~{\rm GeV},~~m_{b}=5.637~{\rm GeV}
\end{eqnarray}
obtained from the fit to heavy baryon spectra. Note that the large
masses of the heavy constituent quarks  in the model of Ref.
\cite{DKT} introduce huge factors $(m_b/m_{b,pole})^5\sim 2.2$ and
$(m_c/m_{c,pole})^5\sim 3.8$ in the semileptonic rate of the heavy
b- and c-hadrons. These factors however  are cancelled to the
great extent by the bound state factor in Eq.
(\ref{w_alpha_beta}).

We display the electron spectra for the inclusive semileptonic
 decays $\Lambda _{b}$ and $\Xi _{b}$ in Fig.~2 and for the inclusive
semileptonic   decays $\Lambda _{c}$ and $\Xi _{c}$ in Fig.~3. The
calculation uses the LF distribution function (\ref{F}) of the
heavy quark derived from the ET D'Souza-Kalman momentum
distribution (\ref{gaussian_distribution}). The perturbative
${\cal O}(\alpha_s)$
 corrections are also included.

The total inclusive semileptonic width $\Gamma (\Lambda_b\to X_c e
\nu_e)=0.086\cdot|V_{cb}/0.04 |^2~{\rm ps}^{-1} $ obtained by a
direct integration of the spectrum coincides with that found in
Ref. \cite{DKKN00}. This rate, though a bit large, agrees with the
estimation based on the semiinclusive branching ratio which (using
$\tau_{\Lambda_b}= 1.229\pm 0.080$ ps$^{-1}$) implies
$\Gamma (\Lambda_b\to X_c e \nu_e)= (7.97^{+2.52}_{-3.09})\times
10^{-2}~ps^{-1}$.
The predicted exclusive/inclusive semileptonic ratio (based on the
estimation  $\Gamma(\Lambda_b\to\Lambda_c)=0.052$~ps$^{-1}$
\cite{KM00}) is
\begin{equation}
R_E(\Lambda_b)=\frac{\Gamma(\Lambda_b\to\Lambda_ce\nu)}{\Gamma(\Lambda_b\to
X_ce\nu)}= 57 \%,
\end{equation}
close to the meson value $R_E({\cal B})=66\%$.

\begin{figure}
\begin{center}
\epsfxsize=7.5cm \epsfysize=7.5cm \epsfbox{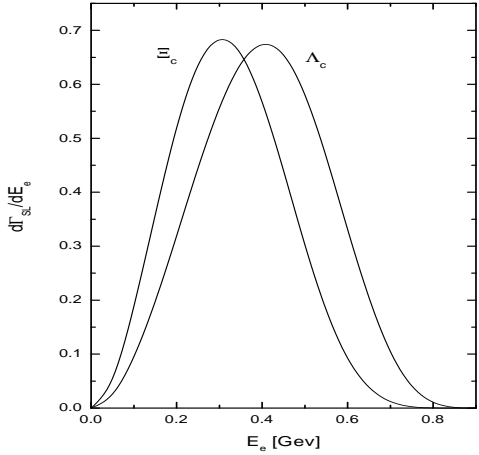}
\end{center}
\vskip -1cm \caption{The predicted electron energy spectrum for $%
\Lambda_c$ and $\Xi_c$ semileptonic decays. The spectra normalized
to $\Gamma_{SL}(\Lambda_c)=0.267$ ps$^{-1}$ and
$\Gamma_{SL}(\Xi_c)=0.236$ ps$^{-1}$}
\end{figure}

The $\Xi _{b}$ inclusive semileptonic rate in the Kalman-D'Souza
model is found to be
$\Gamma (\Xi_{b}\to X_{c}e\nu _{e})=0.092~|V_{cb}/0.04|^{2}~{\rm
ps}^{-1}. $
Note that $\Gamma _{SL}(\Lambda _{b})$ and $\Gamma _{SL}(\Xi
_{b})$ show $\sim 5\%$ flavour dependence due to {\it i}) the
difference of the u and s constituent quark masses and {\it ii})
the difference of $M_{\Lambda _{b}}$ and $M_{\Xi _{b}}$. Both
quantities modify $|\psi (x,p_{\bot }^{2})|^{2}$ marginally.

For the charm decays our result are
$\Gamma (\Lambda _{c} \to X_{s}e\nu _{e})=0.267~{\rm ps}^{-1},$
and  $\Gamma (\Xi _{c} \to X_{s}e\nu _{e})=0.236~{\rm ps}^{-1}.$
The prediction for $\Gamma (\Lambda _{c}\to X_{s}e^{+}\nu )$
compares favorably with the estimation based on the branching
ratio ${\cal BR}(\Lambda _{c}\to e^{+}+anything)=(4.5\pm 1.7)~\%$
and the $\Lambda _{c}$ lifetime $\tau _{\Lambda _{c}}=0.208$ ps:
$\Gamma (\Lambda _{c}\to X_{s}e^{+}\nu )=0.206\pm 0.07~{\rm
ps}^{-1}. $

Using the experimental value $\Gamma(\Lambda_c\to\Lambda_se^+\nu)$
measured by CLEO Collaboration one obtains

\begin{equation}
R_E(\Lambda_c)=\frac{\Gamma(\Lambda_c\to\Lambda_se^+\nu)}{%
\Gamma(\Lambda_c\to X_ce^+\nu)}= (35\pm 8)~ \%.
\end{equation}

\section{Conclusions}
This is the most consistent quark model to-date describing baryons
in the sense that a single set of parameters is used for the whole
spectra and then subjecting all of their properties namely the
masses, radii and their wavefunction structures to a strict test
by a decay model. The spectroscopy is successfully calculated
using far less parameters than any other model \cite{DKT}. Decays
of baryons are calculated. Now in this paper we have successfully
computed the electron spectra in the inclusive semileptonic decays
of the heavy baryons $H_{Q}$ by incorporating the LF formalism.
The existing experimental data are very scarce. There are only a
few upper limits on the total semileptonic widths of heavy
baryons. The further experimental investigation of the
semileptonic decays of heavy flavor baryons will shed the light
upon the validity of OPE in the baryon sector and allows us to
test existing quark models of heavy baryons. \vspace{5mm}

\noindent This work was supported in part RFBR grants Refs.00-02-16363 and
00-15-96786. 

\end{document}